\documentclass[iicol,sn-aps]{sn-jnl}

\jyear{2021}%

\theoremstyle{thmstyleone}%
\theoremstyle{thmstyletwo}%

\theoremstyle{thmstylethree}%

\begin{document}

\title[On Radiative Recombination]{On Quantum and Classical Treatments of Radiative Recombination}


\author[1,2]{\fnm{A. L.} \sur{Barabanov}}

\author[1]{\fnm{K. M.} \sur{Belotsky}}

\author[1]{\fnm{E. A.} \sur{Esipova}}

\author[1]{\fnm{D. S.} \sur{Kalashnikov}}\email{impermast@gmail.com}

\author[1,3]{\fnm{A.~Yu.} \sur{Letunov}}
\equalcont{These authors contributed equally to this work.}


\affil[1]{\orgname{National Research Nuclear University MEPhI}, \orgaddress{\city{Moscow}, \country{Russia}}}

\affil[2]{\orgname{National Research Centre ''Kurchatov Institute''}, \orgaddress{\city{Moscow}, \country{Russia}}}

\affil[3]{\orgdiv{Department}, \orgname{Russian Federal Nuclear Center --- All-Russian Institute of Technical Physics (RFNC--VNIITF)}, \orgaddress{\city{Snezhinsk}, \country{Russia}}}


\abstract{The quantum-mechanical solution to the problem of radiative recombination of an electron in a Coulomb field, obtained in the approximation of the smallness of the electron coupling with the radiation field, has been known for a long time. However, in astrophysics, the classical approach, which does not explicitly use this smallness, is sometimes used to describe similar processes in systems of magnetic monopoles or self-interacting dark matter particles. The importance of such problems is determined by the fact that recombination processes play a crucial role in the evolution of the large-scale structure of the Universe. Therefore, of particular interest is the fact that the classical and quantum expressions for the recombination cross section differ significantly in magnitude. It is shown that the applicability of quantum and classical approaches to radiative recombination is closely related to the radiated angular momentum and its quantization. For situations where the classical approach is not suitable, a semi-classical approach based on the angular momentum quantization is proposed, in some respects an alternative to the well-known semi-classical Kramers' approach.}

\maketitle

\section{Introduction}\label{sec1}

One of the main problems of modern cosmology is the nature of dark matter (DM). Among the hypotheses under consideration is the assumption that DM consists of relatively slow massive particles interacting through Coulomb-type forces. Such particles emit ''dark photons'' in collisions and form ''dark atoms'' in recombination processes. Further, the dark matter ''atoms'' 
 under the influence of gravitational forces form structure involving baryonic matter. Dymanics of structure formation is crucially dependent on DM interaction properties with effectiveness of ''atom'' production \cite{Khlopov2003, Belotsky:2015gsa,BoundStateAnih,DarkAtoms_astro,Cirelli2017,Agrawal_2017,Petraki2017,Khlopov2021,DarkAtom_2022}.
Such models of self-interacting dark matter are now becoming especially popular. They can solve the well-known problems of the standard collisionless scenario of cold dark matter structure formation (for example, the cusp crisis, the predicted excess of dwarf galaxies, the “too big to fail” problem) and suggest ways to search for DM signals in cosmic rays and underground experiments, as well as provide an explanation for the early formation of quasars \cite{Feng_2021}. 

In such models, an important quantity is the radiative recombination (RR) rate of DM particles, which is determined by the recombination cross section. A similar problem for ''usual matter'' related to the radiative recombination of an electron in a Coulomb field is thoroughly studied in atomic physics --- see, e.g., a detailed review by Kotelnikov and Milstein \cite{kotelnikov2019electron} with an extensive bibliography and explanations of the importance of this problem for many applications, including those outside of atomic physics (but without mentioning cosmology). An approximate expression for the cross section for the radiative capture of free electrons into a bound state with the principal quantum number~$n$ was first obtained by Kramers \cite{kramers1923xciii} almost a hundred years ago:
\begin{equation}
\label{1.2}
\sigma^{\rm K}_{n}(\eta)=\frac{32\pi}{3\sqrt{3}}\,\alpha^3a_B^2\,\frac{\eta^4}{n(n^2+\eta^2)}\,.
\end{equation}
Here, $\alpha=e^2/(\hbar c)\simeq 1/137$ is the fine structure constant, $a_B=\hbar^2/(me^2)$ is the Bohr radius, where $m$ is the electron mass, and the dimensionless parameter~$\eta$ is determined by
\begin{equation}
\eta=\frac{Ze^2}{\hbar v},
\label{eta}
\end{equation}
where $Ze$ is the charge of the Coulomb center, and $v$ is the initial velocity of the incident electrons.

It is curious that the Kramers' work~\cite{kramers1923xciii}, published in 1923, before the creating of the consistent quantum theory, was carried out within the framework of the semiclassical approach. However, later in a number of papers, the problem was considered strictly quantum-mechanically, and Eq.~(\ref{1.2}) turned out to be a good approximation to the exact (more cumbersome) result for low-energy electrons (see \cite{kotelnikov2019electron} for details), i.e.,
\begin{equation}
\eta\gg 1.
\label{low_speed_relation}
\end{equation}
In the original work, Kramers considered the radiation of electrons with a sufficiently high energy in the field of the Coulomb center, since he was interested not only in radiative capture, but also in the shape of the high energy part of the bremsstrahlung X-ray spectrum.

The summation of cross sections \eqref{1.2} over all $n$ from 1 to infinity gives an approximate expression for the total RR cross section. Replacing summation by integration introduces an additional minor error, but leads to a simple analytical result
\begin{equation}
\label{1.5}
\sigma^{\rm K}_{RR}(\eta)=\frac{16\pi}{3\sqrt{3}}\,\alpha^3a_B^2\,\eta^2\ln{(1+\eta^2)},
\end{equation}
which is usually called the Kramers formula. In the problem of electron capture by protons for a practically significant region, where $50 <\eta <250$, the Kramers formula slightly overestimates the cross section compared to the result of rigorous quantum mechanical calculations consistent with measurements. However, with a correction factor $\sim 0.9$, it correctly reproduces the dependence of the total RR cross section on the velocity $v$ of incident electrons (see \cite{kotelnikov2019electron} for details).

So, it would seem that the total RR cross section for DM particles is given by Eq.~(\ref{1.5}) with some $\alpha$, $a_B$ and $\eta$, specific for the ''dark matter world''. This is done in many works (see, for example, \cite{Cirelli2017,Petraki2017}), but the situation is not so simple. Indeed, the strict quantum approach leading to (\ref{1.5}) is based on the following approximation: ''The interaction of the electron with the radiation field has been treated as a small pertubation (with the fine structure constant $\alpha$ as the pertubation parameter) and only the lowest order term kept in the expansion in powers of $\alpha$'' \cite{Bethe_1957} (in other words, this is a one-photon approximation). Thus, if the dark matter fine structure constant $\alpha$ is not small,
then the analog of Eq.~(\ref{1.5}) cannot be used as the total RR cross section for the dark matter particles. In this case, some other quantum expression for the RR cross section is needed, obtained for an arbitrary value of $\alpha$ and describing radiative recombination, for example, of a light particle with a charge $z(-e)$ and a heavy particle with a charge $Ze$ without the assumption of smallness of the parameter $z^2e^ 2/(\hbar c)$. However, as far as we know, there is no such formula in the literature.

Thus, the classical expression for the total RR cross section can be useful because, although quantum theory is a true theory describing reality, under certain conditions (in the classical limit) processes are satisfactorily described by classical equations. Such classical expression for the case $z=1$ was apparently first given by Yelutin \cite{elutin}. Let an electron scattered by a stationary Coulomb center with a charge of $Ze$ loses part of its energy for radiation and, as a result, passes into a state of finite motion, i.e., is captured by the Coulomb center. The cross section for such process is given by
\begin{equation}
\sigma_{\rm Cl}(v)=\pi\left(4\pi\right)^{2/5}r^2_0\, Z^{8/5}\left(\frac{c}{v}\right)^{14/5}.
\label{Yelutin}
\end{equation}
Here $r_0=e^2/(mc^2)$ is the classical electron radius. This result is easily generalized to the case $z>1$ (this will be done below).

The potential significance of this result for astrophysical applications is determined by the following. Comparing it with the Kramers formula \eqref{1.5}, rewritten as
\begin{equation}
\sigma^{\rm K}_{RR}(v)=\frac{16\pi}{3\sqrt{3}}\,\alpha r^2_0\left(\frac{Zc}{v}\right)^2 
\ln\left(1+\left(\frac{Ze^2}{\hbar v}\right)^2\right),
\label{Kramers_integral}
\end{equation}
we see, that, firstly, the classical cross section \eqref{Yelutin} has a different dependence on the initial velocity of the electron, and, secondly, the prefactors are very different, mainly due to the absence of the fine structure constant in the classical expression. Both differences lead to the fact that the classical RR cross section significantly exceeds the quantum one, and the excess is stronger the lower the velocity of incident particles. The classical approach to the estimation of RR cross sections was used in the works on determination of concentration of magnetic monopoles \cite{zeldovich1978concentration}\footnote{Though, direct recombination of free monopoles (in contrast to their diffusion process in plasma) was not found to be the main factor determining their relic number density.} and self-interacting dark matter  in the Universe \cite{Belotsky:2005dk,Belotsky:2015fuc,Belotsky:2015rhp,Belotsky_2016,Belotsky2017positron,Nazarova:2017xaw}.

However, the question of the applicability of the expression (\ref{Yelutin}) and its relation to the quantum result (\ref{Kramers_integral}) for the recombination cross section still remains unresolved. In Refs. \cite{elutin,Belotsky_2016,belotsky2020problems}, the condition $\eta\gg Z^{-1}\alpha^{-3/2}$ was obtained, which overlaps with \eqref{low_speed_relation} and corresponds to extremely low incident velocities of the electron for reasonable values of $Z$. In fact, this result cannot be considered satisfactory. Indeed, since the Planck constant $\hbar$ and the dimensionless fine structure constant $\alpha$ are absent in the classical expression (\ref{Yelutin}), there is no way to give it the form (\ref{Kramers_integral}) for at least some interval for the velocity $v$. Therefore, we can assume that in the case of $z=1$, when the parameter $z^2e^2/(\hbar c)$ is small, the classical approach is not valid. Conversely, in the case when the parameter $z^2e^2/(\hbar c)$ is large and the recombination process is not described by the quantum formula (\ref{Kramers_integral}), the classical approach to radiative recombination may be correct, at least in a certain range of particle collision velocities.

Thus, regardless of astrophysical applications, the following question seems to be important: what is the criterion that determines under what parameters of the problem the classical description of the RR cross section is valid, and under which it is not? This question, raised, in particular, in \cite{Belotsky_2016}, seems to us to be fundamental, and this paper is mainly devoted to it.

The method we use is partially borrowed from the classic Kramers' work \cite{kramers1923xciii}. His main idea was to extend the classical consideration of radiative recombination to such a microscopic system as an electron and a nucleus, using the Bohr correspondence principle (subsequently, the same approach was applied to other radiation processes: see, for example, the review \cite{kogan1992kramers} devoted to the so-called Kramers electrodynamics).

As for our work, it is based on a circumstance that seems to have escaped the attention of researchers. Careful consideration based on Maxwell's equations reveals the fundamental reason for the inapplicability of the classical approach to radiative recombination for an electron and a nucleus. It consists in the fact that the change in the angular momentum of an electron cannot be less than the Planck constant $\hbar$ (it is shown below that in the classical approach, the loss of the angular momentum of an electron in the RR process is much less than $\hbar$).

However, combining the classical consideration with the fact of angular momentum quantization, we propose a semi-classical approach, in some sense an alternative to the one used by Kramers. Thus, first, we obtain an expression for the total RR cross section of the same type as the  Kramers formula. Second, we show that nevertheless the classical approach to radiative recombination might be applied to a some class of microscopical systems in astrophysical problems.

\section{Classical Approach}\label{sec2}

Let us first consider the classical RR problem in a rather general formulation. A particle of mass $m$, moves towards the Coulomb center with an initial velocity $v$ and an impact parameter $\rho$. The charges of the Coulomb center and the particle are equal to $Ze$ and $z(-e)$, respectively. If we neglect the effect of radiation on the shape of the trajectory, then the latter is hyperbolic (see, e.g, \cite{landau1988theoretical}). The trajectory equation in the polar coordinates $(r,\varphi)$ can be written in the following form:
\begin{equation}
r(\varphi)=\frac{p}{1+\varepsilon\cos{\varphi}}\,,
\label{r_phi}
\end{equation}
where the orbit parameter $p$ and the eccentricity $\varepsilon$,
\begin{equation}
p=\frac{L^2}{mzZe^2}\,,\quad 
\varepsilon=\sqrt{1+\frac{2L^2E}{m(zZe^2)^2}}\,,
\label{parameters}
\end{equation}
depend on energy $E$ and angular momentum $L$ of the incident particle. The values of these integrals of motion are determined by $v$ and $\rho$: $E=mv^2/2$, $L=m\rho v$. If the direction of the $z$ axis coincides with the vector~$L$, then the azimuthal angle changes from $-\varphi_0$ to $\varphi_0$, where
\begin{equation}
\varphi_0=\arccos\left(-\frac{1}{\varepsilon}\right).
\label{angle}
\end{equation}

Note that the slow incident particles (they are what we are interested in according to Eq.~(\ref{low_speed_relation})), having relatively low energy $E$, move along hyperbolas with eccentricity~$\varepsilon$ close to unity (see Eq.~(\ref{parameters})). Such particles, as it is easy to show, acquire a significant maximum velocity at the point of closest approach to the Coulomb center at $\varphi=0$,
\begin{equation}
v_{\rm max}=v\sqrt{\frac{\varepsilon+1}{\varepsilon-1}}\,.
\label{vmax}
\end{equation}

In the dipole approximation, the radiation losses are determined by the well-known relations \cite{landau1975classical}:
\begin{equation}
\label{2.6}
\frac{dE}{dt}=-\frac{2(ze)^2\,{\ddot{\mathbf{r}}}^{\,2}}{3c^3}\,,\quad
\frac{dL}{dt}=-\frac{2(ze)^2[\dot{\mathbf{r}}\times\, \ddot{\mathbf{r}}\,]_z}{3c^3}\,.
\end{equation}
The total radiation losses $\Delta E=\, -\vert\Delta E\vert$ and $\Delta L=\,-\vert\Delta L\vert$ can be easily found by substituting the acceleration of particle from Newton's law of motion,
\begin{equation}
\label{2.7}
m\,\ddot{\!\mathbf{r}}=-\frac{zZe^2}{r^3}\,\mathbf{r},
\end{equation}
into the above equations. It is convenient to replace $t$ with an azimuthal angle $\varphi$ using the relations
\begin{equation}
\label{2.4}
L=mr^2\dot\varphi
\quad\Rightarrow\quad
dt=\frac{mr^2}{L}\,d\varphi.
\end{equation}
Then, in particular, for the total energy loss we obtain the following expression:
\begin{equation}
\label{2.9}
\vert \Delta E\vert =\frac{4m(z^3Z^2e^5)^2}{3c^3L^5}\,f(\varphi_0),
\end{equation}
where
\begin{equation}
\label{2.9.2}
f(\varphi_0)=\varphi_0\left(1+\frac{1}{2\cos^2\varphi_0}\right)-\frac{3}{2}\tan\varphi_0.
\end{equation}
Similar formulas, but for $z=1$, are presented, in particular, in the original Kramers' paper \cite{kramers1923xciii}.

Obviously, the main contribution to the radiation comes from the part of the trajectory where the particle is in the closest position to the Coulomb center. Since the energy loss due to radiation $\vert\Delta E\vert$ is small compared to the characteristic kinetic energy $mv_{\rm max}^2/2$ of the particle in this region, the effect of radiation on the shape of this part of the trajectory can be neglected. If, however, the initial energy of the particle $E=mv^2/2$ is so small, that
\begin{equation}
\label{2.11.1}
\vert\Delta E\vert>E,
\end{equation}
the particle passes from a hyperbolic orbit to a highly elongated elliptical one, i.e. it is captured by the Coulomb center. This is the effect of radiative recombination in the classical approach.

Thus, for a given initial energy $E=mv^2/2$ the effect takes place for sufficiently large energy loss \eqref{2.9}, i.e. for small angular momenta $L$ and, hence, impact parameters $\rho$. However, there are limitations for $L$ (and $\rho$) from below due to following reason. Equations \eqref{2.6} are valid in all parts of the trajectory, where the dipole approximation is applicable, i.e. the particle velocity is much smaller than the speed of light. Obviously, this is the case when $v_{\rm max}\ll c$, that leads to the following restriction for the angular momentum:
\begin{equation}
L=m\rho v\gg\frac{(1+\varepsilon)zZe^2}{c}\,.
\label{2.4.4}
\end{equation}
It means, that trajectories with low values of the impact parameter fall out of consideration. However, as it will be shown below, it does not affect the validity of the Eq.~(\ref{Yelutin}).

For angular momenta values, which simultaneously satisfy \eqref{2.4.4} and small enough to provide high energy losses $\vert\Delta E\vert$, one obtains: $\varepsilon\to 1$ and $\varphi_0\to \pi$. So, accordingly \eqref{2.9} and \eqref{2.9.2}, the energy loss for radiation is equal to
\begin{equation}
\label{EnergyLossFromLL}
\vert\Delta E\vert=\frac{2\pi m(z^3Z^2e^5)^2}{c^3L^5}\,,
\end{equation}
see similar Eq.~(23) in \cite{kramers1923xciii} and Eq.~(7) in \cite{elutin}, both obtained for $z=1$.

\par Now, the condition (\ref{2.11.1}) can be rewritten as follows 
\begin{equation}
\label{2.11.2}
(4\pi)^{2/5}(z^3Z^2)^{4/5}r_0^2\left(\frac{c}{v}\right)^{14/5}>\rho^2\,.
\end{equation}
The left-hand side of the relation \eqref{2.11.2} can be denoted as $\rho_m^2$. It means that the condition of the transition into the bound state (for the Coulomb potential) can be formulated as: $\rho<\rho_m$. Thus, the classical total RR cross section is equal to $\pi\rho_m^2$. Substitution of $z=1$ leads to the expression \eqref{Yelutin}. Note, that the condition \eqref{2.4.4} for $\rho=\rho_m$ turns into  $v\ll c\sqrt{z/Z}$, which is valid for moderate values of $z$ and $Z$ (because the incident particles in our treatment are non-relativistic).

Let us now consider particles with such small impact parameters that the condition \eqref{2.4.4} is not satisfied. Thus, these particles achieve relativistic velocities near the Coulomb center. However, radiation losses for such particles are much greater than $\vert\Delta E\vert$ for non-relativistic particles. Therefore, the condition \eqref{2.11.1} is satisfied for these particles too, and they also turn into bound states.

\section{Kramers' Approach}

Within the framework of the classical approach described above, each particle with a given initial velocity $v$ and an impact parameter $\rho$ emits a strictly defined energy $\vert\Delta E\vert$, and goes into a bound state if the energy satisfies the condition \eqref{2.11.1}. The situation is different in the semiclassical approach. Here, the particle passes into a bound state only with a certain probability $P(\rho)$ (below, we assume the initial velocity $v$ to be fixed, but trace the dependence of the quantities on $\rho$). The corresponding semiclassical cross section is equal to
\begin{equation}
\label{crosssection}
\sigma=2\pi\int\limits_{0}^{\infty}P(\rho)\,\rho d\rho.
\end{equation}

Kramers' semiclassical approach (developed for electrons, respectively, $z=1$ in this section) is based on the spectral decomposition of energy loss into radiation: 
\begin{equation}
\label{spectrum}
\vert\Delta E(\rho)\vert=\int\limits_{0}^{\infty}I(\rho,\omega)d\omega.
\end{equation}
From a naive quantum point of view, the energy flux emitted by electrons (with given initial velocity $v$ and impact parameter $\rho$) in the frequency range from $\omega$ to $\omega+d\omega$ is provided by photons with energy~$\hbar\omega$. Let us assume that an electron with probability $q (\rho,\omega)d\omega$ emits a photon with the energy~$\hbar\omega$. Then according to the correspondence principle we equate the total energy $NI(\rho,\omega)d\omega$, emitted by $N$ electrons in the frequency range from $\omega$ to $\omega+d\omega $ in the classical approach, and the total energy $\hbar\omega Nq(\rho,\omega)d\omega\,$ emitted by the same $N$ electrons in the same frequency range in the semiclassical approach. This gives the relation
\begin{equation}
\label{difprobability}
q(\rho,\omega)=\frac{I(\rho,\omega)}{\hbar\omega}\,.
\end{equation}

If then, following Kramers \cite{kramers1923xciii}, we match to each bound state with the energy $E_n=-Z^2e^2/(2a_Bn^2)$ a finite frequency interval $\Delta_n\omega$, we obtain for the probability of an electron transition to a state with the principal quantum number~$n$ the following expression:
\begin{equation}
P_n^{K}(\rho)=\int\limits_{\Delta_n\omega}q(\rho,\omega)d\omega.
\label{nprobability}
\end{equation}
This transition is accompanied by the formation of a discreet line in the bremsstrahlung spectrum corresponding to the photon energy
\begin{equation}
\hbar\omega_n=\frac{mv^2}{2}+\vert E_n\vert.
\label{omegan}
\end{equation}
Calculating the integral \eqref{crosssection} with the probabilities (\ref{nprobability}) leads to the formula \eqref{1.2}.

Note that the continuous part of the bremsstrahlung spectrum, described by effective radiation (see \cite{landau1975classical})
\begin{equation}
\kappa(\omega)=2\pi\int\limits_{0}^{\infty}q(\rho,\omega)\hbar\omega\rho d\rho
\,\,\,\equiv 2\pi\int\limits_{0}^{\infty}I(\rho,\omega)\rho d\rho,
\label{kappa}
\end{equation}
covers the following photons energies,
\begin{equation}
\hbar\omega<\frac{mv^2}{2}\equiv\hbar\omega_0,
\label{omegas}
\end{equation}
where $\omega_0$ is the boundary frequency. The high energy part of this spectrum adjoining $\omega_0$ is ''flat'', i.e. is proportional to $\theta(\omega_0-\omega)$. Using an explicit expression for the classical differential spectrum $I(\rho,\omega)$, Kramers obtained this ''flat shape'', 
\begin{equation}
\kappa(\omega)=\frac{16\pi Z^2e^6}{3\sqrt{3}\,c^3m^2v^2}\,.
\label{kappaflat}
\end{equation}
This result is reproduced, in particular, in Ref. \cite{landau1975classical}, see Eq. (70.22).

However, the Kramers' semiclassical approach gives us no idea about the area of applicability of the classical expression \eqref{Yelutin} for the total RR cross section.

\section{Alternative Semiclassical Approach}

That is why we return to the classical approach and note that the energy loss \eqref{EnergyLossFromLL}, which is necessary for binding the system, is inevitably accompanied by an angular momentum loss. The latter is determined in accordance with Eqs. \eqref{2.6}, \eqref{2.7} by the following relation:
\begin{equation}
\frac{dL}{dt}=-\dfrac{2z^3Ze^4}{3m^2c^3r^3}L.
\label{brem_an_mom}
\end{equation}
Using \eqref{2.4} to replace the time $t$ by the azimuthal angle~$\varphi$ as well as the expressions \eqref{r_phi}, \eqref{parameters} and the result $\varphi_0\simeq\pi$, we obtain for the total loss of angular momentum
\begin{equation}
\vert\Delta L\vert=\frac{4\pi (z^2Ze^3)^2}{3L^2c^3}\,,
\label{ddl}
\end{equation}
with the use of additional assumption $\vert\Delta L\vert\ll L$, which validity is ensured by the condition \eqref{2.4.4}.

Then we take into account the fundamental fact that angular momentum is discrete
\begin{equation}
L=l\hbar,
\label{q_ang_mom}
\end{equation}
where $l$ is an integer number. It means that \eqref{ddl} can be rewritten in the form
\begin{equation}
\vert\Delta L\vert=\frac{4\pi (z^2Z)^2}{3l^2}\left(\frac{e^2}{\hbar c}\right)^3\hbar.
\label{ddl2}
\end{equation}
It is easy to see from \eqref{ddl2} that $\vert\Delta L\vert\ll \hbar$, if $z$ and $Z$ are not too high. It is for this reason that the classical approach is inapplicable to the problem of describing radiation capture in such an elementary system as an electron and a nucleus.

However, this opens up the possibility of constructing a semiclassical approach different from Kramers' method. For generality, consider the radiative recombination in the system of two particles with masses $m_1$, $m_2$  and charges $Z_0e$, $z_0(-e)$, respectively, with an interaction $U(r)=-z_0Z_0e^2/r$, where $r$ --- distance between the particles. This problem reduces to the equivalent problem with a particle with the reduced mass $m=m_1m_2/(m_1+m_2)$ and the charge $z(-e)$ on the stationary Coulomb center with the charge $Ze$, where $z=(z_0m_1+Z_0m_2)/(m_1+m_2)$ and $Z=z_0Z_0/z$.

Suppose that a particle (with an initial velocity~$v$ and an impact parameter $\rho$) radiates only with probability $P(\rho)$, but loses (in the one act) the angular momentum~$\hbar$, as well as an energy $\vert\Delta E^{\prime}\vert$. Then, taking into account the correspondence principle, and equating angular momentum $N\vert\Delta L\vert$, emitted by $N$ particles in the classical approach, and the same value $NP(\rho)\,\hbar$ in the semiclassical approach, we get
\begin{equation}
P(\rho)=\frac{\vert\Delta L\vert}{\hbar}=
\frac{4\pi(z^2Ze^3)^2}{3\hbar m^2\rho^2v^2c^3}\,.
\label{probability2}
\end{equation}
Doing the same for energy losses, $N\vert\Delta E\vert$ and $NP(\rho)\vert\Delta E^{\prime}\vert$, we find
\begin{equation}
\vert\Delta E^{\prime}\vert\equiv\hbar\omega(\rho)=\frac{\vert\Delta E\vert}{P(\rho)}=
\frac{3\hbar (zZe^2)^2}{2m^2\rho^3v^3}\,,
\label{energy2}
\end{equation}
where the radiated energy is equated to the energy of photon with the frequency $\omega(\rho)$\footnote{The frequency $\omega(\rho)$ differs from the maximum angular frequency $\omega_{\rm max}$ of an electron moving in a parabolic orbit at the point of closest approach to the Coulomb center by a scale factor close to one. It was pointed out in the review~\cite{KoganPlanckConstant} that the characteristic frequency of bremsstrahlung radiation is also close to $\omega_{\rm max}$, and the relation $\hbar\omega_{\rm max}\gg\,\vert\Delta E\vert$ means that, in the semiclassical approach, the bremsstrahlung radiation is the result of strong fluctuations.}. Thus, in this approach the differential spectrum takes the form
\begin{equation}
I^{\prime}(\rho,\omega)=P(\rho)\,\hbar\omega(\rho)\,\delta(\omega-\omega(\rho)).
\label{spectrumprime}
\end{equation}
In principle, one can introduce a probability $q^{\prime}(\rho,\omega)=P(\rho)\delta(\omega-\omega(\rho))$ related with $I^{\prime}(\rho,\omega)$ in the same way that $q(\rho,\omega)$ is related with $I(\rho,\omega)$ by Eq. (\ref{difprobability}). Then the quantity $P(\rho)$ (\ref{probability2}) can be considered as the probability integrated over frequencies $\omega$.

Thus, we obtain a simple description of the entire integral (averaged over $\rho$) spectrum of the bremsstrahlung radiation, continuous, with photon energies (\ref{omegas}), and discrete, caused by the radiative recombination, with photon energies lying in the interval:
\begin{equation}
    \hbar\omega_0<\hbar\omega(\rho)<\hbar\omega_{max}=\frac{mv^2}{2}+\frac{z^2Z^2e^2}{2a_B}\,.
    \label{emit_cond}
\end{equation}
Indeed, the effective radiation (\ref{kappa}) with the differential spectrum (\ref{spectrumprime}) reproduces the expected ''flat'' high energy part of the bremsstrahlung radiation,
\begin{equation}
\kappa^{\prime}(\omega)=2\pi P(\rho)\,\hbar\omega(\rho)\,\rho\, \biggl\vert \frac{d\rho}{d\omega}\biggr\vert =\frac{8\pi^2z^4Z^2e^6}{9c^3m^2v^2}\,,
\label{kappa2}
\end{equation}
for $\omega$ close to $\omega_0$, as the Kramers' method do. On the other hand, using the relation (\ref{energy2}) between $\omega$ and $\rho$, we get from \eqref{emit_cond} the interval for corresponding impact parameters,
\begin{equation}
\rho_{min}=\frac{\hbar}{mv}\sqrt[3]{\frac{3(z\eta)^2}{1+(z\eta)^2}}
<\rho<
\rho_0=\frac{\hbar\sqrt[3]{3(z\eta)^2}}{mv}.
\label{rho_cond}
\end{equation}
Then, substituting the energy \eqref{energy2} into the left-hand side of Eq.~\eqref{2.11.1} leads to the right inequality of Eq.~\eqref{rho_cond}. It means that the total RR cross section is given by the integral \eqref{crosssection} with the probability $P(\rho)$ \eqref{probability2} and the limits of integration determined by Eq.~\eqref{rho_cond}. The expression \eqref{crosssection} contains the integral
\begin{equation}
\int\limits_{\rho_{\rm min}}^{\rho_{0}}\frac{d\rho}{\rho}=
\ln\left(\frac{\rho_{0}}{\rho_{\rm min}}\right)=
\dfrac{1}{3}\ln{\bigg(1+(z\eta)^2 \bigg)}.
\label{integral}
\end{equation}
The total RR cross section takes the form
\begin{equation}
\sigma_{RR}^{(z)}(\eta)=\frac{8\pi^2}{9}\alpha^3a_B^2\,z^2 (z\eta)^2\ln (1+(z\eta)^2).
\label{final_sigma}
\end{equation}

This expression at $z=1$ and the Kramers formula \eqref{1.5} differ only in prefactors. Their ratio is equal to
\begin{equation}
    \dfrac{8\pi^2/9}{16\pi/(3\sqrt{3})}\approx 0.9,
\end{equation}
i.e. close to unity. Note that the relationship between expressions \eqref{kappa2} at $z=1$ and \eqref{kappaflat} for the continuous ''flat spectrum'', obtained in the framework of Kramers and the presented semiclassical approaches, is exactly the same.

So, using the condition
\begin{equation}
{\vert\Delta L\vert\ll \hbar,}
\label{4.1}
\end{equation}
under which the classical theory is obviously wrong, we derived, within the framework of our semiclassical approach, the Kramers formula (with slightly different numerical coefficient), consistent with exact quantum mechanical calculations. Substituting Eq. (\ref{ddl}) into the left-hand side of (\ref{4.1}), we get the range of impact parameters,
\begin{equation}
{\sqrt{\frac{4\pi}{3\alpha}}\,z^2\eta\, r_0\ll \rho.}
\label{4.1.2}
\end{equation}
Obviously, the impact parameters satisfying the inequalities (\ref{rho_cond}) must also lie in this range, i.e.
\begin{equation}
{\sqrt{\frac{4\pi}{3\alpha}}\,z^2\eta r_0\ll \rho_{\rm min}<\frac{\sqrt[3]{3}\,\hbar}{mv}.}
\label{4.1.3}
\end{equation}
This gives the condition
\begin{equation}
{z^2Z\alpha^{3/2}\ll \sqrt[3]{3}\,\sqrt{\frac{3}{4\pi}}\simeq 0.7,}
\label{4.1.4}
\end{equation}
that is always true in the world of ordinary elementary objects, like electrons and atomic nuclei, with $\alpha\simeq 1/137$.

However, in the dark matter world this may not be the case. Therefore, consider the inverse conditions
\begin{equation}
{\vert\Delta L\vert\gg \hbar,\qquad \sqrt{\frac{4\pi}{3\alpha}}\,z^2\eta\, r_0\gg \rho,}
\label{4.2}
\end{equation}
corresponding, apparently, to the classical limit ($\hbar\to 0$). In accordance with Eq. (\ref{probability2}), the value of $P(\rho)\gg 1$ now takes the meaning of the number of emitted photons. Indeed, a photon emerging in the process of dipole radiation carries away an angular momentum equal to $\hbar$. Thus, the classical description corresponds to the multiphoton case. The total radiated energy turns out to be equal $\Delta E=P(\rho)\hbar\omega(\rho)$ (see Eq. (\ref{energy2})), defined by the classical expression (\ref{EnergyLossFromLL}). Thus, radiative recombination occurs if the previously written conditions (\ref{2.11.1}) and (\ref{2.11.2}) are met, so that the total RR cross section takes the form
\begin{equation}
{\sigma_{Cl}^{(z)}(v)=\pi\rho_m^2=
\pi (4\pi)^{2/5}(z^3Z^2)^{4/5}r_0^2\left(\frac{c}{v}\right)^{14/5},}
\label{4.3}
\end{equation}
coinciding with (\ref{Yelutin}) at $z=1$.

But now we can write down the condition for the applicability of the classical expression for the total RR cross section. This follows from the requirement that the impact parameters $\rho\le\rho_m$ lie in the range defined by the second of the inequalities (\ref{4.2}), i.e.
\begin{equation}
{\rho_m\ll \sqrt{\frac{4\pi}{3\alpha}}\,z^2\eta\, r_0.}
\label{4.4}
\end{equation}
This leads to the following condition on the velocity of the incident particles:
\begin{equation}
{\frac{v}{c}\gg\frac{3^{5/4}}{(4\pi)^{3/4}z^2\sqrt{Z}\alpha^{5/4}}\simeq
\frac{0.6}{z^2\sqrt{Z}\alpha^{5/4}}\,.}
\label{4.5}
\end{equation}
We see that if the dark matter fine structure constant $\alpha=e^2/(\hbar c)$ is not small and at least one of the parameters $z$ and $Z$ significantly exceeds unity, then there exists a region of relatively low velocities in which the total RR cross section is described by the classical formula (\ref{4.3}). Recall, in particular, that magnetic monopoles have a charge $q_M\sim \hbar c/e$.

\section{Conclusion}

Classical and quantum, based on the smallness of fine structure constant, approaches to the description of radiative recombination in a system of two particles bound by an attractive Coulomb interaction lead to significantly different expressions for the cross sections. Therefore, it is important to understand under what conditions each of these approaches is valid.

In this paper, we draw attention to the importance of taking into account the magnitude of the angular momentum of radiation emitted during recombination. In the electron-nucleus system, this value, calculated within the framework of classical electrodynamics, always turns out to be significantly less than the Planck constant $\hbar$. Accordingly, the change in the angular momentum of an electron passing into a bound state turns out to be the same. But this is impossible from a quantum point of view, since the angular momentum of an electron can only change by a multiple of $\hbar$. Therefore, in this situation, as we showed, the total RR cross section is correctly described by the rigorous quantum or both semiclassical (proposed by Kramers and by authors of this work) approaches, but not by the classical one. The resulting condition (\ref{4.1.4}) for the validity of the Kramers formula is essentially the requirement that the fine structure constant be small.

However, in systems of magnetic monopoles or self-interacting dark matter, the analogue of the fine structure constant can be significant. Then the Kramers formula is wrong, and due to the lack of an exact quantum formula for this case, we can use the classical formula in its range of applicability (\ref{4.5}). Thus, for the problem of radiative recombination in a system of nonrelativistic particles, a lower bound on their velocity has been obtained. Obviously, only in the case when the fine structure constant is not small, this restriction can be satisfied in a certain range of values of the parameters $z$ and $Z$. 

The cases of large charges and, possibly, the emission of many photons with very low energies require non-perturbative treatment, which deserves separate consideration.
\bigskip

\bmhead{Acknowledgments}

We would like to thank the referees for useful comments and criticism, S.G.Rubin for interest to this work with useful discussion and A.B.Kukushkin for valuable consultations on the literature devoted to the problems considered, also Yu.B.Gurov, M.P.Faifman, M.D.Skorokhvatov and M.N.Alfimov for assistance of different kind in solving the problem. The work of A.L.B. (on elaborating solution of the problem within semiclassical approach) was supported by the MEPhI Program Priority 2030. The work of K.M.B. (on connection of this task with dark matter problem) was supported by Ministry of Science and Higher Education of the Russian Federation by project No 0723-2020-0040 “Fundamental problems of cosmic rays and dark matter”. The work of E.A.E. (on calculation of applicability conditions) was supported by fund Basis, project № 18-1-5-89-1.

\end{document}